\documentclass[runningheads]{llncs}
\usepackage{graphicx}
\usepackage{url}
\usepackage{float}
\usepackage{amsmath}
\usepackage{subfig}
\usepackage{dblfloatfix}
\usepackage[table]{xcolor}
\usepackage{color, colortbl}
\definecolor{Gray}{gray}{0.9}
\usepackage{graphicx,dblfloatfix}
\usepackage{tikz}
\usepackage{lipsum,afterpage}
\begin{document}

\mainmatter  


\title{Empirical Analysis on Effectiveness of NLP Methods for Predicting Code Smell}
\titlerunning{NLP Methods for predicting code smell}

\author{Himanshu Gupta\thanks{The research associated to this paper was completed during author's undergraduate study at BITS Pilani, Hyderabad Campus.} 
\and Abhiram Anand Gulanikar*
\and Lov Kumar  \and
Lalita Bhanu Murthy Neti 
}
\authorrunning{H. Gupta et al.}

\institute{
	BITS Pilani, Hyderabad Campus, India 
	\email{ \{f20150339h,f20150105h\}@alumni.bits-pilani.ac.in}
	\email{ \{lovkumar,bhanu\}@hyderabad.bits-pilani.ac.in} 
}

\toctitle{Lecture Notes in Computer Science}
\tocauthor{Authors' Instructions}

\maketitle
\begin{abstract}

A code smell is a surface indicator of an inherent problem in the system, most often due to deviation from standard coding practices on the developer’s part during the development phase. Studies observe that code smells made the code more susceptible to call for modifications and corrections than code that did not contain code smells. Restructuring the code at the early stage of development saves the exponentially increasing amount of effort it would require to address the issues stemming from the presence of these code smells. Instead of using traditional features to detect code smells, we use user comments (given on the packages’ repositories) to manually construct features to predict code smells. We use three Extreme learning machine kernels over 629 packages to identify eight code smells by leveraging feature engineering aspects and using sampling techniques. Our findings indicate that the radial basis functional kernel performs best out of the three kernel methods with a mean accuracy of 98.52.

\end{abstract}

\section{Introduction}
The existence of code smells in source code points towards poor design and violations in standard coding practices\cite{van2002java}. The code smells may not necessarily be identified as defects in the software in the current phase, but these code classes have a high likelihood of developing bugs in the future. Since these code smells do not cause defects, the only way to identify them is based on inspection, i.e., manually combing through thousands of lines of code to find code smells. This method is highly disorganized and costly and becomes more inefficient along with scaling of code package size. In our work, we are automating this process of identifying code smells. We are using the input source code packages to build our set of source code metrics to develop a model to locate and predict code smells in the application packages. These models will reduce the cost and efficiency of maintaining software while enforcing standard coding practices and improving its quality.

In this paper, we used three kernels: Linear kernel, radial basis function kernel, and polynomial kernel to develop models to predict the following eight code smells, namely, Swiss Army Knife (SAK), Long Method (LM), Member Ignoring Method (MIM), No Low Memory Resolver (NLMR), Blob Class (BLOB), Internal Getter/Setter (IGS), Leaking Inner Class (LIC) and Complex Class (CC). These source code metrics are from the application’s source code packages and are used to engineer relevant features and select relevant metrics. We used the Wilcoxon Rank Sum Test to achieve the second of these objectives. In this work, we have analyzed the performance with various kernel functions using accuracy, area under the curve (AUC), and F-measure to predict code smells.We have attempted to answer three research questions in this paper:

\begin{itemize}
    \item \textbf{RQ1: Discuss the ability of different NLP Methods to generate features that help detect code smells.} In traditional code smell detection techniques, code smell metrics are present, which help detect code smells. We have manually constructed 100 features derived from reviews of peer developers’ software about the software’s source code in this problem. We have a Continuous Bag of words and the Skip-gram method to construct features for detection. We will use accuracy, Area under the curve, and F1 Score to compare each technique’s performance. 
    
    \item \textbf{RQ2: Explore the potential of Data Sampling Techniques to discover code smells} Instead of using just the original data, we have used three sampling techniques to generate datasets. SMOTE\cite{chawla2002smote} (Synthetic Minority Over-sampling), borderline SMOTE\cite{han2005borderline}, and SVM SMOTE (Support Vector Machine SMOTE)\cite{mathew2015kernel}, along with original data, gives us four sets of data. We compare the performance of these datasets using Area under curve and statistical significance tests.

    \item \textbf{RQ3: Study the capacity of various ELM Kernels to predict code smells.} We have used three Extreme Learning Machine kernels to detect code smells from the various features and datasets. Linear Kernel (LINK), Radial Basis Function kernel (RBF), and Polynomial kernel have been used for classification. Their performance has been compared using statistical significance tests and Area Under the Curve Analysis.
    
 \end{itemize}   
    
\noindent \textbf{Organization:} The paper is prepared as follows: The 2\textsuperscript{nd} section summarizes the associated work. The 3\textsuperscript{rd} section offers an in-depth review of all the components used in the experiment. The 4\textsuperscript{th} section  describes the study's framework pipeline and how the components described in section 3 interact with each other.The 5\textsuperscript{th} section provides the experimental outcome and the 6\textsuperscript{th} section answers the questions raised in the introduction. In the 7\textsuperscript{th} section we conclude our research.

\section{Related Work}
Evgeniy et al. used contextual analysis of document data to generate features making use of word sense clarification. Long Ma et al. used Word2Vec to output word vectors to represent large pieces of texts or entire documents. He used CBOW and skip-grams as component models of Word2Vec to create word vectors and then evaluate word similarity.\cite{ma2015using} Hui Han et al. introduced over-sampling techniques of Borderline-SMOTE as a variant of SMOTE where only minority examples near borderline are over-sampled. \cite{han2005borderline} Josey Mathew et al. proposed a kernel-based SMOTE(SVM SMOTE)  algorithm which directly generates the minority data points. His proposed SMOTE technique performs better than other SMOTE techniques in 51 benchmark datasets.\cite{mathew2015kernel} Guang-Bin Huang et al. proposed Extreme Learning Machine(ELM), which randomly chooses hidden nodes and determines the Single-hidden Layer Feed forward Neural Networks(SLFN) weights. \cite{huang2006extreme} Francisco Fernandez-Navarro et al. proposed a modified version of ELM, which uses Gaussian distribution to parameterize the distribution called the radial basis function. She used ELM is used to optimize the parameters of the model.\cite{fernandez2011melm}
\vspace{-0.5cm}

\begin{table}[b!]
\centering
\vspace{-0.5cm}
	\caption{Statistics on code smell distribution by type}
	\resizebox{10cm}{!}
        {
            \begin{tabular}{|l|l|l|l|l|}
\hline
\textbf{\begin{tabular}[c]{@{}l@{}}Code smell \\ names\end{tabular}} & \textbf{\begin{tabular}[c]{@{}l@{}}Repository \# without \\ any code smell\end{tabular}} & \textbf{\begin{tabular}[c]{@{}l@{}}Repository \# with\\ code smell\end{tabular}} & \textbf{\begin{tabular}[c]{@{}l@{}}Percent of classes\\ without code smell\end{tabular}} & \textbf{\begin{tabular}[c]{@{}l@{}}Percent of classes\\ with code smell\end{tabular}} \\ \hline
\textbf{Complex Class (CC)}                                          & 230                                                                                     & 399                                                                             & 63.4\%                                                                                    & 36.5\%                                                                                 \\ \hline
\textbf{Leaking Internal Class (LIC)}                                & 160                                                                                     & 469                                                                             & 74.5\%                                                                                    & 25.4\%                                                                                 \\ \hline
\textbf{Blob Class}                                                  & 460                                                                                     & 169                                                                             & 26.8\%                                                                                    & 73.1\%                                                                                 \\ \hline
\textbf{No Low Memory Resolver (NLMR)}                               & 190                                                                                     & 439                                                                             & 69.7\%                                                                                    & 30.2\%                                                                                 \\ \hline
\textbf{Internal Getter Setter (IGS)}                                & 264                                                                                     & 365                                                                             & 58.02\%                                                                                   & 41.9\%                                                                                 \\ \hline
\textbf{Member Ignoring Method (MIM)}                                & 265                                                                                     & 364                                                                             & 57.8\%                                                                                    & 42.1\%                                                                                 \\ \hline
\textbf{Swiss Army Knife (SAK)}                                      & 155                                                                                     & 474                                                                             & 75.3\%                                                                                    & 24.6\%                                                                                 \\ \hline
\textbf{Long method (LM)}                                            & 225                                                                                     & 404                                                                             & 64.2\%                                                                                    & 35.7\%                                                                                 \\ \hline 
\end{tabular}
        }
\end{table}

\section{Research Methodology}

\begin{figure*}[t!]
	\centering
	\includegraphics[width=12cm, height= 4cm]{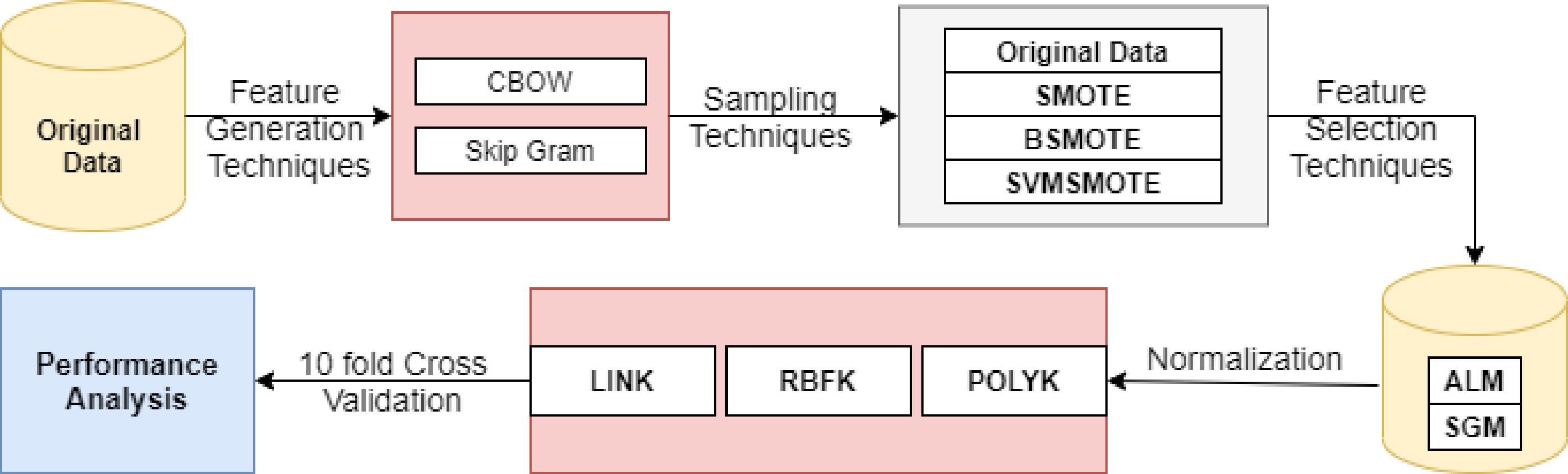}
	\caption{Flowchart of the Research Framework}
	\label{fig1}
	\vspace{-0.5cm}
\end{figure*}

A detailed description of the Dataset, Data Sampling Techniques, Feature Generation Techniques, Feature Selection Techniques, and Classification Algorithms is given below.

\vspace{-0.5cm}
\subsection{Experimental Data Set}

In this research, our main database comprised of 629 freely available software packages. Our dataset consisted of a list of packages, and the code smells present in them. The characteristics and patterns exhibited by all of the code smells are presented in Table 1. Table 1 shows that the code smells are present in a range from 25.4\% to 73.1\%. We also observed that the lowest presence of any code smell we found was 25.4\% for the BLOB Class code smell, while the highest presence observed at 75.04\% was for Swiss Army Knife (SAK) code smell.

\vspace{-0.5cm}
\subsection{Data Sampling Techniques}

We use three data sampling techniques to generate additional datasets to mitigate the bias in the dataset :
\begin{itemize}
    \item \textbf{SMOTE}\cite{chawla2002smote} randomly chooses samples from K nearest neighbors from minority class. The synthetic data would be made between the randomly selected sample and the k nearest neighbor.
    \item \textbf{Borderline SMOTE}\cite{han2005borderline} works on a similar principle but creates data only along the classes’ boundary line, not to introduce external bias.
    \item \textbf{SVM-SMOTE}\cite{wang2017novel,mathew2015kernel} uses Support Vector Machine instead of K nearest neighbor to generate samples between a chosen sample and the decision boundary.
\end{itemize}
\vspace{-0.5cm}
\subsection{Feature Generation Techniques}
We use two architectures from the word2vec \cite{ma2015using} techniques, namely the Continuous Bag of Words (CBOW) and Skip-gram. Continuous Bag of Words method \cite{wang2017two} uses the surrounding words to speculate the present word. Since it derives from the bag of words model, words present in the window( surrounding the current word) are not differentiated based on the current word’s distance. The skip-gram model \cite{guthrie2006closer} makes use of the current word to predict the context. Here words nearer are more heavily weighted than words farther away from the present word. Comparing the two models, the CBOW model is faster than skip-grams, but skip-grams perform better when uncommon words are involved.


\vspace{-0.5cm}
\subsection{Feature Selection Techniques}
We have generated 100 feature metrics, but they might not be relevant to the code smells we have considered. We use the Wilcoxon signed-rank test to get the statistical relation between the smelly and clean applications. We have set 0.05 as the outset for the p-value, and we reject the hypothesis if the value is lower. We employ cross-correlation analysis to select uncorrelated features. Our selected variables share a high correlation to the output variables and have a low correlation between themselves.
\vspace{-0.5cm}

\begin{figure*}[t!]
	\centering
	\includegraphics[width=11cm, height= 4.5cm]{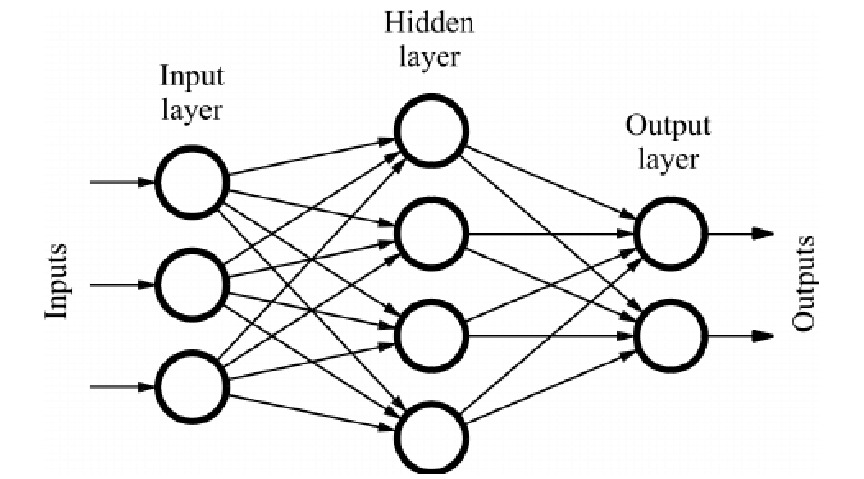}
	\caption{Architecture diagram of Extreme Machine Learning Kernel}
	\label{fig2}
	\vspace{-0.5cm}
\end{figure*}

\subsection{Classification Algorithms}

\vspace{-0.25cm}

This paper uses three ELM kernel functions \cite{huang2006extreme,micchelli2005learning} to train models to predict code smells, namely the Linear Kernel function, Radial basis kernel function, and polynomial kernel function\cite{prajapati2010performing}.As shown in Figure 2, Extreme Learning Machines (ELM) can be simply defined as feed-forward neural networks, and they can be used for clustering, classification, regressing, among other things. These three kernel methods work best for different data types based on whether it is linearly separable and the problem is linear or nonlinear.  Kernel functions are mathematical functions used to transform training data into higher dimensions.
The linear kernel is generally chosen when dealing with linearly separable data; it is also most commonly used when many features are in a dataset.
The Radial basis function kernel is a non-linear kernel used for training SVMS when solving nonlinear problems.
The polynomial kernel function is also used to train nonlinear models.
It is faster and requires fewer resources to train the linear or polynomial kernel functions than radial basis functions. Still, they are less accurate in comparison to the RBF kernel.
We also use ten-fold cross-validation to overcome overfitting and selection-bias issues and obtain insights on our model’s performance on an independent dataset. We use the area under the curve (AUC) and F-measure, among other tests, to compare their performance.
\vspace{-0.5cm}
\section{Research Framework}

\vspace{-0.25cm}

We make use of the code data from 629 open-source software packages on GitHub. To eliminate the class imbalance problem in the data, we use SMOTE, Borderline SMOTE, and SVM SMOTE \cite{chawla2002smote,mathew2015kernel,han2005borderline}to get four datasets: the Original Dataset (ORD) SMOTE Dataset, Borderline SMOTE Dataset, and SVM SMOTE Dataset. We use three kernel functions, the linear kernel function, the radial basis kernel function, and the polynomial kernel function. To compare the accuracy over all the four datasets, we have created and used the area under the curve (AUC) and F-measure, among other tests, to compare their performance. Figure 1 provides a clear representation of the same.

\begin{table*}[t!]
	\renewcommand{\thesubfigure}{\thefigure.\arabic{subfigure}}
	\centering
	\vspace{-0.5cm}
	\caption{Area Under Curve and Accuracy figures for ELM models trained on the original dataset}
	\label{table1}
	\subfloat[Accuracy values\label{tab1s1}]
	{
		\renewcommand{\arraystretch}{1.1}
		\resizebox{10cm}{!}
        {
        \begin{tabular}{|l|r|r|r|r|r|r|l|r|r|r|r|r|r|}
        \hline
        \multicolumn{14}{|c|}{\textbf{Original Data}}                                                                                                                                                                                                                                                                                                                                                                                                                                       \\ \hline
         & \multicolumn{6}{c|}{\textbf{ALM}}                                                                                                                                                                                             &  & \multicolumn{6}{c|}{\textbf{SGM}}                                                                                                                                                                                             \\ \cline{2-14} 
                          & \multicolumn{3}{c|}{\textbf{CBOW}}                                                                            & \multicolumn{3}{c|}{\textbf{SKG}}                                                                             &  & \multicolumn{3}{c|}{\textbf{CBOW}}                                                                            & \multicolumn{3}{c|}{\textbf{SKG}}                                                                             \\ \cline{2-14} 
                          & \multicolumn{1}{c|}{\textbf{LINK}} & \multicolumn{1}{c|}{\textbf{RBFK}} & \multicolumn{1}{c|}{\textbf{POLYK}} & \multicolumn{1}{c|}{\textbf{LINK}} & \multicolumn{1}{c|}{\textbf{RBFK}} & \multicolumn{1}{c|}{\textbf{POLYK}} &  & \multicolumn{1}{c|}{\textbf{LINK}} & \multicolumn{1}{c|}{\textbf{RBFK}} & \multicolumn{1}{c|}{\textbf{POLYK}} & \multicolumn{1}{c|}{\textbf{LINK}} & \multicolumn{1}{c|}{\textbf{RBFK}} & \multicolumn{1}{c|}{\textbf{POLYK}} \\ \hline
        \textbf{BLOB}     & 62.80                              & 100.00                             & 100.00                              & 73.77                              & 100.00                             & 100.00                              &  & 62.64                              & 100.00                             & 72.02                               & 63.91                              & 68.04                              & 76.63                               \\ \hline
        \textbf{LM}       & 75.36                              & 100.00                             & 100.00                              & 75.68                              & 99.84                              & 100.00                              &  & 75.36                              & 100.00                             & 100.00                              & 75.36                              & 100.00                             & 100.00                              \\ \hline
        \textbf{SAK}      & 75.20                              & 100.00                             & 100.00                              & 74.72                              & 100.00                             & 100.00                              &  & 75.83                              & 75.83                              & 100.00                              & 73.77                              & 100.00                             & 94.28                               \\ \hline
        \textbf{CC}       & 71.70                              & 100.00                             & 100.00                              & 71.54                              & 100.00                             & 100.00                              &  & 70.27                              & 100.00                             & 100.00                              & 70.75                              & 100.00                             & 93.80                               \\ \hline
        \textbf{IGS}      & 67.73                              & 99.68                              & 100.00                              & 68.04                              & 63.43                              & 100.00                              &  & 61.84                              & 100.00                             & 65.66                               & 57.71                              & 74.56                              & 100.00                              \\ \hline
        \textbf{MIM}      & 58.98                              & 100.00                             & 100.00                              & 59.46                              & 82.35                              & 100.00                              &  & 59.62                              & 100.00                             & 94.59                               & 59.62                              & 88.71                              & 100.00                              \\ \hline
        \textbf{NLMR}     & 81.08                              & 100.00                             & 100.00                              & 81.56                              & 100.00                             & 100.00                              &  & 75.04                              & 100.00                             & 99.84                               & 75.04                              & 94.75                              & 93.16                               \\ \hline
        \textbf{LIC}      & 70.11                              & 100.00                             & 100.00                              & 65.18                              & 85.06                              & 100.00                              &  & 64.86                              & 100.00                             & 97.46                               & 66.45                              & 78.54                              & 100.00                              \\ \hline
        \end{tabular}
        }
	}

	\subfloat[AUC Values\label{tab1s2}]
	{
		\renewcommand{\arraystretch}{1.1}
		\resizebox{10cm}{!}
        {
            \begin{tabular}{|l|r|r|r|r|r|r|l|r|r|r|r|r|r|}
\hline
\multicolumn{14}{|c|}{\textbf{Original Data}}                                                                                                                                                                                                                                                                                                                                                                                                                                        \\ \hline
 & \multicolumn{6}{c|}{\textbf{ALM}}                                                                                                                                                                                             &  & \multicolumn{6}{c|}{\textbf{SGM}}                                                                                                                                                                                             \\ \cline{2-14} 
                  & \multicolumn{3}{c|}{\textbf{CBOW}}                                                                            & \multicolumn{3}{c|}{\textbf{SKG}}                                                                             &  & \multicolumn{3}{c|}{\textbf{CBOW}}                                                                            & \multicolumn{3}{c|}{\textbf{SKG}}                                                                             \\ \cline{2-14} 
                  & \multicolumn{1}{c|}{\textbf{LINK}} & \multicolumn{1}{c|}{\textbf{RBFK}} & \multicolumn{1}{c|}{\textbf{POLYK}} & \multicolumn{1}{c|}{\textbf{LINK}} & \multicolumn{1}{c|}{\textbf{RBFK}} & \multicolumn{1}{c|}{\textbf{POLYK}} &  & \multicolumn{1}{c|}{\textbf{LINK}} & \multicolumn{1}{c|}{\textbf{RBFK}} & \multicolumn{1}{c|}{\textbf{POLYK}} & \multicolumn{1}{c|}{\textbf{LINK}} & \multicolumn{1}{c|}{\textbf{RBFK}} & \multicolumn{1}{c|}{\textbf{POLYK}} \\ \hline
\textbf{BLOB}     & 0.64                               & 1.00                               & 1.00                                & 0.78                               & 1.00                               & 1.00                                &  & 0.60                               & 1.00                               & 0.78                                & 0.62                               & 0.86                               & 0.84                                \\ \hline
\textbf{LM}       & 0.72                               & 1.00                               & 1.00                                & 0.75                               & 1.00                               & 1.00                                &  & 0.72                               & 1.00                               & 1.00                                & 0.67                               & 1.00                               & 1.00                                \\ \hline
\textbf{SAK}      & 0.71                               & 1.00                               & 1.00                                & 0.68                               & 1.00                               & 1.00                                &  & 0.69                               & 0.72                               & 1.00                                & 0.65                               & 1.00                               & 0.99                                \\ \hline
\textbf{CC}       & 0.75                               & 1.00                               & 1.00                                & 0.70                               & 1.00                               & 1.00                                &  & 0.65                               & 1.00                               & 1.00                                & 0.69                               & 1.00                               & 0.98                                \\ \hline
\textbf{IGS}      & 0.74                               & 1.00                               & 1.00                                & 0.74                               & 0.74                               & 1.00                                &  & 0.65                               & 1.00                               & 0.73                                & 0.61                               & 0.84                               & 1.00                                \\ \hline
\textbf{MIM}      & 0.63                               & 1.00                               & 1.00                                & 0.61                               & 0.91                               & 1.00                                &  & 0.62                               & 1.00                               & 0.99                                & 0.61                               & 0.96                               & 1.00                                \\ \hline
\textbf{NLMR}     & 0.84                               & 1.00                               & 1.00                                & 0.85                               & 1.00                               & 1.00                                &  & 0.71                               & 1.00                               & 1.00                                & 0.67                               & 0.99                               & 0.98                                \\ \hline
\textbf{LIC}      & 0.76                               & 1.00                               & 1.00                                & 0.67                               & 0.98                               & 1.00                                &  & 0.67                               & 1.00                               & 1.00                                & 0.68                               & 0.94                               & 1.00                                \\ \hline
\end{tabular}
        }
	} 
	\vspace{-0.5cm}
\end{table*}

\begin{figure*}[b!]
    \vspace{-0.5cm}
	\centering
	\includegraphics[width=12cm, height= 4cm]{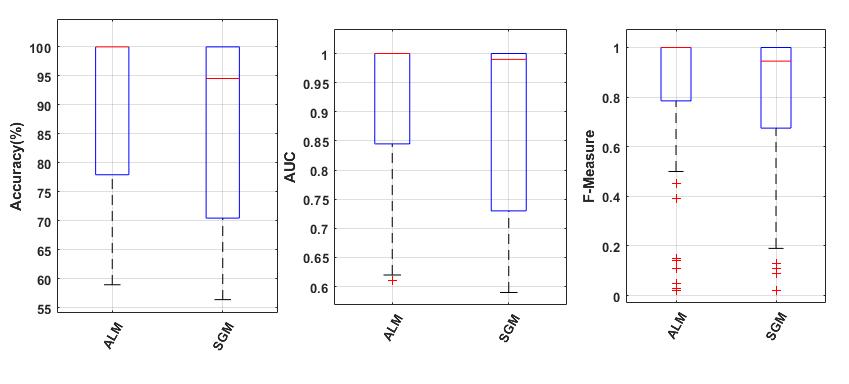}
	\vspace{-0.5cm}
	\caption{Box plot comparison between All Metrics}
	\label{fig3}
	
\end{figure*}

\section{Experimental Results}

Tables 2A and 2B give accuracy and AUC values for all the ELM methods, using feature engineering methods and all feature selection techniques. Table 3 and Table 4 summarize the various statistical measures of different metrics used in our research. It is pretty evident from Tables 2A and 2B that Radial Basis Function and polynomial perform much better for most samples than Linear kernel. Table 3B Figure 3 shows that models trained using all metrics perform better than those using significant metrics. The high values of the performance indicators encourage the use of the code smell prediction model. The following observations can be made made from the Results obtained :

\begin{itemize}
    \item The performance of all models varies greatly with the minimum accuracy being 56.41\% and the maximum accuracy being 100\% AUC follows similar trend to accuracy but f-measure varies the most from minimum value being 0.02 and maximum being 1.
    \item It is observed that radial basis kernel and polynomial kernel perform much better than linear kernel across all three statistical measures that is AUC, accuracy and f-measure and they also indicate the high efficiency of the models which are developed.
    \item It is observed that Linear kernel performs the best with Class NLMR (77.5\%) and the worst with MIM class. (61.95\%)
\end{itemize}

\vspace{-0.5cm}

\begin{table*}[t!]
    \vspace{-0.5cm}
	\renewcommand{\thesubfigure}{\thefigure.\arabic{subfigure}}
	\centering
	\caption{Statistical Measure}
	\label{table2}
	\subfloat[Feature Generation Techniques\label{tab2s1}]
	{
		\renewcommand{\arraystretch}{1.1}
		\resizebox{6.5cm}{!}
		{
			\begin{tabular}{|l|r|r|r|r|r|r|}
\hline
              & \multicolumn{1}{l|}{\textbf{Min}} & \multicolumn{1}{l|}{\textbf{Max}} & \multicolumn{1}{l|}{\textbf{Mean}} & \multicolumn{1}{l|}{\textbf{Median}} & \multicolumn{1}{l|}{\textbf{25th}} & \multicolumn{1}{l|}{\textbf{75th}} \\ \hline
\multicolumn{7}{|c|}{\textbf{Accuracy}}                                                                                                                                                                                                     \\ \hline
\textbf{CBOW} & 57.98                             & 100.00                            & 88.63                              & 100.00                               & 74.61                              & 100.00                             \\ \hline
\textbf{SKM}  & 56.41                             & 100.00                            & 87.82                              & 99.85                                & 73.98                              & 100.00                             \\ \hline
\multicolumn{7}{|c|}{\textbf{AUC}}                                                                                                                                                                                                          \\ \hline
\textbf{CBOW} & 0.59                              & 1.00                              & 0.90                               & 1.00                                 & 0.78                               & 1.00                               \\ \hline
\textbf{SKM}  & 0.59                              & 1.00                              & 0.90                               & 1.00                                 & 0.80                               & 1.00                               \\ \hline
\multicolumn{7}{|c|}{\textbf{F Measure}}                                                                                                                                                                                                    \\ \hline
\textbf{CBOW} & 0.02                              & 1.00                              & 0.86                               & 1.00                                 & 0.74                               & 1.00                               \\ \hline
\textbf{SKM}  & 0.05                              & 1.00                              & 0.84                               & 1.00                                 & 0.72                               & 1.00                               \\ \hline
\end{tabular}
	    }
    } 
    \subfloat[Features Selection Metrics\label{tab2s2}]
	{
		\renewcommand{\arraystretch}{1.1}
		\resizebox{6.5cm}{!}
		{
			\begin{tabular}{|l|r|r|r|r|r|r|}
\hline
             & \multicolumn{1}{l|}{\textbf{Min}} & \multicolumn{1}{l|}{\textbf{Max}} & \multicolumn{1}{l|}{\textbf{Mean}} & \multicolumn{1}{l|}{\textbf{Median}} & \multicolumn{1}{l|}{\textbf{25th}} & \multicolumn{1}{l|}{\textbf{75th}} \\ \hline
\multicolumn{7}{|c|}{\textbf{Accuracy}}                                                                                                                                                                                                    \\ \hline
\textbf{ALM} & 58.98                             & 100.00                            & 90.72                              & 100.00                               & 77.94                              & 100.00                             \\ \hline
\textbf{SGM} & 56.41                             & 100.00                            & 85.73                              & 94.55                                & 70.46                              & 100.00                             \\ \hline
\multicolumn{7}{|c|}{\textbf{AUC}}                                                                                                                                                                                                         \\ \hline
\textbf{ALM} & 0.61                              & 1.00                              & 0.93                               & 1.00                                 & 0.84                               & 1.00                               \\ \hline
\textbf{SGM} & 0.59                              & 1.00                              & 0.88                               & 0.99                                 & 0.73                               & 1.00                               \\ \hline
\multicolumn{7}{|c|}{\textbf{F Measure}}                                                                                                                                                                                                   \\ \hline
\textbf{ALM} & 0.02                              & 1.00                              & 0.88                               & 1.00                                 & 0.78                               & 1.00                               \\ \hline
\textbf{SGM} & 0.02                              & 1.00                              & 0.83                               & 0.95                                 & 0.68                               & 1.00                               \\ \hline
\end{tabular}
		}
	}
	\vspace{-0.5cm}
\end{table*}

\begin{figure*}[b!]
\vspace{-0.5cm}
	\centering
	\includegraphics[width=12cm, height= 4cm]{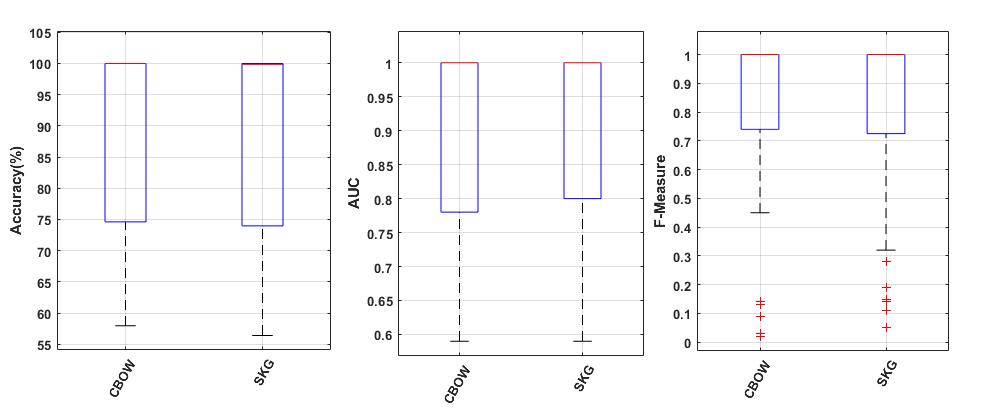}
	\vspace{-0.5cm}
	\caption{Accuracy, AUC and F-measure box-plot of different feature generation techniques}
	\label{fig4}
	
\end{figure*}
\vspace{-0.25cm}
\section{Comparison}
\textbf{RQ1: Discuss the ability of different NLP Methods to generate features that help detect code smells.}\\
Table 3A and Figure 4 shows that CBOW performs slightly better than skip-grams across accuracy and F-measure metric. It is a known fact that CBOW is many times faster to train compared to skip-grams. CBOW performs marginally better when common words are considered, while skip-gram performs better on rare words or phrases. \cite{wang2017novel} Our model performs better on CBOW, indicating that user comments from which the feature vectors are generated have a higher occurrence of common words over rare words.
Table 6B shows us the result of the Ranksum test of vectors generated using these two methods, and we can conclude that the vectors generated are highly uncorrelated.\\

\begin{table*}[t!]
    \vspace{-0.5cm}
	\renewcommand{\thesubfigure}{\thefigure.\arabic{subfigure}}
	\centering
	\caption{Statistics about different datasets and machine learning models}
	\label{table3}
	\subfloat[Performance on various ELM Kernels \label{tab3s1}]
	{
		\renewcommand{\arraystretch}{1.1}
		\resizebox{6.5cm}{!}
		{
			\begin{tabular}{|l|r|r|r|r|r|r|}
\hline
\multicolumn{7}{|c|}{\textbf{Accuracy}}                                                                                                                                                                                                      \\ \hline
\textbf{}      & \multicolumn{1}{l|}{\textbf{Min}} & \multicolumn{1}{l|}{\textbf{Max}} & \multicolumn{1}{l|}{\textbf{Mean}} & \multicolumn{1}{l|}{\textbf{Median}} & \multicolumn{1}{l|}{\textbf{25th}} & \multicolumn{1}{l|}{\textbf{75th}} \\ \hline
\textbf{LINK}  & 56.41                             & 82.59                             & 69.84                              & 70.57                                & 64.39                              & 75.11                              \\ \hline
\textbf{RBFK}  & 63.43                             & 100.00                            & 98.52                              & 100.00                               & 100.00                             & 100.00                             \\ \hline
\textbf{POLYK} & 64.92                             & 100.00                            & 96.32                              & 100.00                               & 95.57                              & 100.00                             \\ \hline
\multicolumn{7}{|c|}{\textbf{AUC}}                                                                                                                                                                                                           \\ \hline
               & \multicolumn{1}{l|}{\textbf{Min}} & \multicolumn{1}{l|}{\textbf{Max}} & \multicolumn{1}{l|}{\textbf{Mean}} & \multicolumn{1}{l|}{\textbf{Median}} & \multicolumn{1}{l|}{\textbf{25th}} & \multicolumn{1}{l|}{\textbf{75th}} \\ \hline
\textbf{LINK}  & 0.59                              & 0.90                              & 0.74                               & 0.74                                 & 0.66                               & 0.80                               \\ \hline
\textbf{RBFK}  & 0.72                              & 1.00                              & 0.99                               & 1.00                                 & 1.00                               & 1.00                               \\ \hline
\textbf{POLYK} & 0.71                              & 1.00                              & 0.98                               & 1.00                                 & 0.99                               & 1.00                               \\ \hline
\multicolumn{7}{|c|}{\textbf{F Measure}}                                                                                                                                                                                                     \\ \hline
               & \multicolumn{1}{l|}{\textbf{Min}} & \multicolumn{1}{l|}{\textbf{Max}} & \multicolumn{1}{l|}{\textbf{Mean}} & \multicolumn{1}{l|}{\textbf{Median}} & \multicolumn{1}{l|}{\textbf{25th}} & \multicolumn{1}{l|}{\textbf{75th}} \\ \hline
\textbf{LINK}  & 0.02                              & 0.86                              & 0.62                               & 0.69                                 & 0.59                               & 0.75                               \\ \hline
\textbf{RBFK}  & 0.32                              & 1.00                              & 0.98                               & 1.00                                 & 1.00                               & 1.00                               \\ \hline
\textbf{POLYK} & 0.45                              & 1.00                              & 0.96                               & 1.00                                 & 0.96                               & 1.00                               \\ \hline
\end{tabular}
	    }
    } 
    \subfloat[Performance on different Datasets\label{tab3s2}]
	{
		\renewcommand{\arraystretch}{1.1}
		\resizebox{6.5cm}{!}
		{
			\begin{tabular}{|l|r|r|r|r|r|r|}
\hline
\textbf{}         & \multicolumn{1}{l|}{\textbf{Min}} & \multicolumn{1}{l|}{\textbf{Max}} & \multicolumn{1}{l|}{\textbf{Mean}} & \multicolumn{1}{l|}{\textbf{Median}} & \multicolumn{1}{l|}{\textbf{25th}} & \multicolumn{1}{l|}{\textbf{75th}} \\ \hline
\multicolumn{7}{|c|}{\textbf{Accuracy}}                                                                                                                                                                                                         \\ \hline
\textbf{ORD}      & 57.71                             & 100.00                            & 86.66                              & 96.10                                & 73.77                              & 100.00                             \\ \hline
\textbf{SMOTE}    & 57.02                             & 100.00                            & 88.78                              & 100.00                               & 74.46                              & 100.00                             \\ \hline
\textbf{BSMOTE}   & 56.41                             & 100.00                            & 88.50                              & 100.00                               & 73.14                              & 100.00                             \\ \hline
\textbf{SVMSMOTE} & 58.69                             & 100.00                            & 88.96                              & 99.95                                & 74.61                              & 100.00                             \\ \hline
\multicolumn{7}{|c|}{\textbf{AUC}}                                                                                                                                                                                                              \\ \hline
\textbf{ORD}      & 0.60                              & 1.00                              & 0.88                               & 0.99                                 & 0.72                               & 1.00                               \\ \hline
\textbf{SMOTE}    & 0.61                              & 1.00                              & 0.91                               & 1.00                                 & 0.81                               & 1.00                               \\ \hline
\textbf{BSMOTE}   & 0.59                              & 1.00                              & 0.91                               & 1.00                                 & 0.80                               & 1.00                               \\ \hline
\textbf{SVMSMOTE} & 0.61                              & 1.00                              & 0.92                               & 1.00                                 & 0.82                               & 1.00                               \\ \hline
\multicolumn{7}{|c|}{\textbf{F Measure}}                                                                                                                                                                                                        \\ \hline
\textbf{ORD}      & 0.02                              & 1.00                              & 0.74                               & 0.96                                 & 0.50                               & 1.00                               \\ \hline
\textbf{SMOTE}    & 0.58                              & 1.00                              & 0.89                               & 1.00                                 & 0.75                               & 1.00                               \\ \hline
\textbf{BSMOTE}   & 0.57                              & 1.00                              & 0.89                               & 1.00                                 & 0.74                               & 1.00                               \\ \hline
\textbf{SVMSMOTE} & 0.59                              & 1.00                              & 0.89                               & 1.00                                 & 0.75                               & 1.00                               \\ \hline
\end{tabular}
		}
	}
	\vspace{-0.5cm}
\end{table*}

\begin{figure*}[b!]
\vspace{-0.5cm}
	\centering
	\includegraphics[width=12cm, height= 4cm]{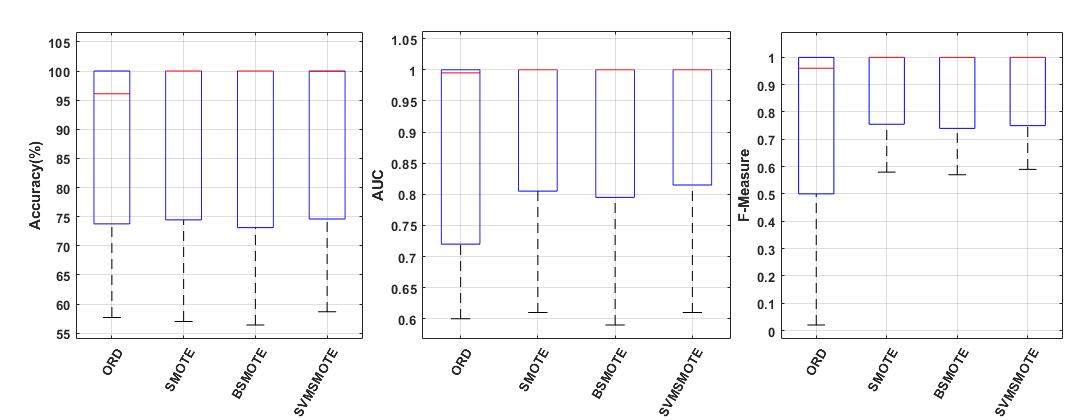}
	\caption{Comparison between different data sampling techniques.}
	\label{fig6}
	\vspace{-0.5cm}
\end{figure*}

\textbf{RQ2: Explore the potential of Data Sampling Techniques to discover code smells.}\\
Table 4B and Figure 5 shows that the data sampling techniques perform better than the original data in AUC, accuracy, and F-measure. Although all three SMOTE techniques, BorderlineSMOTE and SVM-SMOTE, perform nearly the same, SVM-SMOTE performs the best. SVM-SMOTE performs better than others because they use KNN. SVM can employ kernels to lead to a better hyperplane in higher dimensions. KNN uses euclidean distance, on the other hand, which may not work well in the same case. Also, KNN computes the nearest neighbors’ distance, leading to more unsatisfactory performance when working on a large dataset. 

Table 5A gives us the result of the Ranksum test of the datasets generated using these methods. We observe that all the datasets generated from smoothing techniques vary a lot from the original dataset, and we can conclude that the datasets are highly uncorrelated. We also observe that SMOTE, Borderline-SMOTE, and SVM-SMOTE are very similar to each other, and hence the performance of the models trained over them also show similar trends.

\begin{figure*}[t!]
	\centering
	\vspace{-0.5cm}
	\includegraphics[width=12cm, height= 4cm]{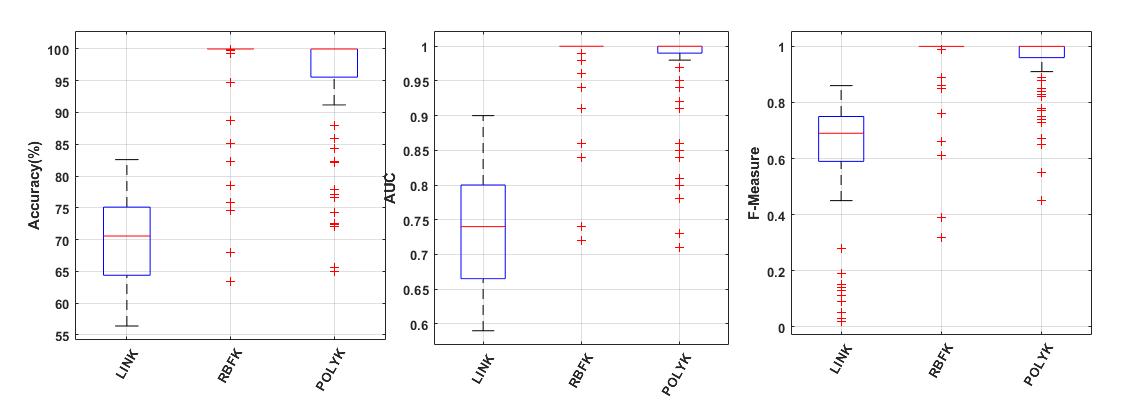}
	\vspace{-0.5cm}
	\caption{Box-plot comparison between different ELM Kernel methods}
	\vspace{-0.5cm}
	\label{fig5}
\end{figure*}

\begin{table*}[t!]

	\renewcommand{\thesubfigure}{\thefigure.\arabic{subfigure}}
	\centering
	\caption{Ranksum Test}
	\vspace{-0.5cm}
	\label{table4}
	\subfloat[Different sampling methods \label{tab4s1}]
	{
		\renewcommand{\arraystretch}{1.1}
		\resizebox{7 cm}{!}
		{
			\begin{tabular}{|l|r|r|r|r|}
\hline
                  & \multicolumn{1}{l|}{\textbf{ORD}} & \multicolumn{1}{l|}{\textbf{SMOTE}} & \multicolumn{1}{l|}{\textbf{BSMOTE}} & \multicolumn{1}{l|}{\textbf{SVMSMOTE}} \\ \hline
\textbf{ORD}      & 1.00                              & 0.12                                & 0.19                                 & 0.12                                   \\ \hline
\textbf{SMOTE}    & 0.12                              & 1.00                                & 0.76                                 & 0.89                                   \\ \hline
\textbf{BSMOTE}   & 0.19                              & 0.76                                & 1.00                                 & 0.83                                   \\ \hline
\textbf{SVMSMOTE} & 0.12                              & 0.89                                & 0.83                                 & 1.00                                   \\ \hline
\end{tabular}
	    }
    } 
    \subfloat[Model similarity \label{tab4s2}]
	{
		\renewcommand{\arraystretch}{1.1}
		\resizebox{4.5cm}{!}
		{
			\begin{tabular}{|l|r|r|r|}
\hline
               & \multicolumn{1}{l|}{\textbf{LINK}} & \multicolumn{1}{l|}{\textbf{RBFK}} & \multicolumn{1}{l|}{\textbf{POLYK}} \\ \hline
\textbf{LINK}  & 1.00                               & 0.00                               & 0.00                                \\ \hline
\textbf{RBFK}  & 0.00                               & 1.00                               & 0.00                                \\ \hline
\textbf{POLYK} & 0.00                               & 0.00                               & 1.00                                \\ \hline
\end{tabular}
		}
	}
	\vspace{-0.5cm}
\end{table*}

\begin{table*}[b!]
\vspace{-0.5cm}
	\renewcommand{\thesubfigure}{\thefigure.\arabic{subfigure}}
	\centering
	\caption{Ranksum Test}
	\label{table5}
	\subfloat[Feature Combination \label{tab5s1}]
	{
		\renewcommand{\arraystretch}{1.1}
		\resizebox{2.5 cm}{!}
		{
			\begin{tabular}{|l|r|r|}
\hline
             & \multicolumn{1}{l|}{\textbf{ALM}} & \multicolumn{1}{l|}{\textbf{SGM}} \\ \hline
\textbf{ALM} & 1.00                              & 0.00                              \\ \hline
\textbf{SGM} & 0.00                              & 1.00                              \\ \hline
\end{tabular}
        }
	}
	\hspace{1cm}
	\subfloat[Feature Generation Methods\label{tab5s2}]
	{
		\renewcommand{\arraystretch}{1.1}
		\resizebox{3 cm}{!}
		{
			\begin{tabular}{|l|r|r|}
\hline
              & \multicolumn{1}{l|}{\textbf{CBOW}} & \multicolumn{1}{l|}{\textbf{SKM}} \\ \hline
\textbf{CBOW} & 1.00                               & 0.43                              \\ \hline
\textbf{SKM}  & 0.43                               & 1.00                              \\ \hline
\end{tabular}
		}
	}
	\vspace{-0.5cm}
\end{table*}

\textbf{RQ3: Study the capacity of various ELM Kernels to predict code smells.}

Table 4A and Figure 6 shows the three kernel methods’ performance in terms of accuracy, AUC, and F-measure. Since our data does not have a linear distribution, we observe that the linear kernel method’s performance is relatively lackluster. Polynomial and RBF both perform significantly better than the linear kernel due to a fixed small number of features. It is observed that the RBF kernel shows the best performance of the three.
Table 5B shows the result of the Ranksum tests on models generated using the different ELM kernels. We can observe that the prediction models developed using the various methods are significantly different from each other, and the models are highly unrelated.

\section{Conclusion}
This paper provides the empirical evaluation of code smell prediction utilizing various ELM methods, feature generation methods using NLP techniques, feature selection, and data sampling techniques. The models are evaluated using ten-fold cross-validation, and their prediction abilities are compared using accuracy, AUC, and F-measure. We draw the following conclusions from our research study:

\begin{itemize}
    \item CBOW performs better than skip-grams in feature generation.
    \item SVM-SMOTE performs best among the data sampling techniques.
    \item Models based on all metrics perform better than models based on significant metrics created using the Wilcoxon signed-rank test.
    \item RBF kernel performs best among the EML methods in predicting code smells.
\end{itemize}

\bibliographystyle{unsrt}
\bibliography{paper}

\end{document}